\documentclass{article}
\usepackage[T1]{fontenc}
\usepackage[utf8]{inputenc}
\usepackage{ismir}
\usepackage{amsmath,cite,url}
\usepackage{graphicx}
\usepackage{color}
\usepackage{xurl}
\usepackage{enumitem}
\setitemize{noitemsep,topsep=0pt,parsep=0pt,partopsep=0pt,leftmargin=*}

\title{Detection of AI-generated stems within \\hybrid human-AI music}

\multauthor
  {Fran\c cois Rigaud \hspace{0.5cm} Gabriel Meseguer-Brocal \hspace{0.5cm} Benjamin Martin \hspace{0.5cm} Romain Hennequin}
  {Deezer Research, Paris, France\\
  {\tt\small research@deezer.com}}

\def\authorname{F. Rigaud \textit{et al.}}

\begin{document}

\maketitle

\begin{abstract}

This paper presents, to the best of our knowledge, the first study on detecting human-AI hybrid music tracks created by mixing human-produced and AI-generated stems. 
Building on recent work showing that AI music detectors can identify decoder-related artifacts in fully generated music, we investigate whether such artifacts remain detectable at the stem level after mixing.
Using MUSDB18-HQ database in a two-stem vocals + accompaniment setting, we simulate hybrid mixtures by autoencoding individual stems with a neural codec. 
We compare two strategies combining AI-generated mix detection and source separation. A naive sequential pipeline, where source separation is followed by detection on separated sources, confirms that artifacts associated with an AI-generated stem are not reliably recovered by generic source separation systems. We therefore propose a parallel architecture in which source separation is only used to estimate source-relative energy within the mixture. We then train simple stem-specific binary classifiers that take as input the generated mix prediction together with the relative energy of the target stem on short audio chunks. Averaging chunk-level predictions yields encouraging track-level results, highlighting the potential of such approaches for detecting AI-generated stems in hybrid music.

\end{abstract}

\section{Introduction}\label{sec:introduction}

%Paragraph 1: Full AI gen and detection
In the past few years, music AI generation has become widespread, with numerous commercial services enabling users to create a full music song from a simple text prompt. Beyond concerns of copyright infringement related to model training and memorization~\cite{yin2022measuring}, the rapid adoption of these services resulted in the flooding of music streaming catalogs, associated for a significant part with fraudulent usage aimed at diverting revenues from actual artists~\cite{mbw2026DeezerFlood}.
One response to this challenge is automated detection, \textit{i.e.} flagging whether a given track has been fully AI-generated.
Recent work has shown that AI-generated music can be detected with high accuracy, even in large-scale industrial contexts~\cite{afchar2025detecting, afchar2025fourier}. These systems exploit systematic artifacts introduced by the transposed convolution layers present in the neural decoders common to most generative models.

% Paragraph 2: Hybrid music production & detection
Early AI music services were mostly limited to generate complete pieces from text prompts, yielding what we call \emph{fully generated AI music}. Modern services, however, offer much finer control over the generated material. These platforms provide music producers with Digital Audio Workstation-like interfaces to interact with generative algorithms and produce individual layers of arrangement from an initial composition. Professional musicians and producers are increasingly incorporating such tools into their creative and production workflows, with major music labels partnering with generative AI services~\cite{wmg_suno2025, umg_stability2025, wmg_stability2025}. 
In this context, \emph{hybrid music}, for which some human-produced stems are mixed together with AI-generated stems, is expected to become increasingly common. 
This raises an important and unaddressed question: \textit{can one distinguish between fully generated AI music and hybrid music?} 
Going beyond binary classification at the global level of a track, and detecting which stems are generated is of importance, as it may affect editorial policies, AI transparency, and artistic authenticity~\cite{Kinzer2024}.
To the authors' knowledge, no scientific methodology has been published on this task.
Note, that in this work, we restrict our definition of hybrid music to the audio recording level. Other configurations, such as human-written lyrics paired with a generated recording, or conversely, generated lyrics paired with a human recording~\cite{frohmann2025double}, fall outside the scope of this study. 

% Paragraph 3: Context of this work
In this work, we focus on the limited two-stem setting of vocals and accompaniment mixtures, which still covers prominent hybrid scenarios enabled by modern generative platforms: AI-generated vocals imitating a specific voice layered over a real accompaniment, or conversely, an AI-generated accompaniment paired with a real vocal recording. 
We formalize the problem as a binary classification task for each stem independently, predicting whether it is AI-generated or not.
While results may vary for other stem types and mixing settings, in the following we provide insights into the factors likely to influence performance beyond these limitations and leave formal generalization to more stems for future work.

% Paragraph 4: Contributions and plan
This paper presents three main contributions.
First, we demonstrate experimentally in Section~\ref{sec:fakemix_globaldetector} that state-of-the-art binary detectors fail to reliably handle hybrid mixes: depending on the source energy ratio, they either miss the generated stem or incorrectly flag the entire track.
Second, in Section~\ref{sec:naive_source_separation_detector}, we evaluate a naive approach applying source separation prior to detection. We find that artifacts from generated stems are not reliably isolated by generic separation systems and spread over all separated stems, making this pipeline insufficient on its own.
Third, building on insights from these experiments, we propose in Section~\ref{sec:hybrid_detection_approach} lightweight stem-specific classifiers that take as input local mix-level detection scores and per-band source energy ratios. We show that aggregating chunk-level predictions yields substantially better stem-level performance than the naive approach.
Together with this article, we provide a code repository\footnote{github.com/deezer/ismir26-hybrid-human-ai-music-detection} for reproducing all the experiments.

\section{Related work}\label{sec:related_work}

\textbf{AI-Generated Music Detection:}
AI-generated music detection has attracted increasing interest in recent years, with most approaches relying on black-box deep learning models that offer little understanding of their decision criteria~\cite{dauban2024ircam,afchar2025detecting,crosvila2025arms,rahman2024sonics}. Alternatively, checkerboard artifacts arising from transposed convolutions in the decoder have been identified as a reliable and interpretable detection signal~\cite{afchar2025fourier}. These artifacts are a strong clue for discriminating generated music from genuine one. They are architecture-dependent and highly detectable, allowing a reliable binary classification that is consistently able to detect various models, including commercial ones~\cite{afchar2025detecting, afchar2025fourier}. 
An alternative paradigm frames detection as audio tagging, \textit{e.g.} using CLAP embeddings~\cite{dauban2024ircam, crosvila2025arms}. 
All these approaches show significant vulnerability and drop in performance to transformations such as sample rate conversion, pitch shifting, or time stretching~\cite{crosvila2025arms, afchar2025detecting, afchar2025fourier}. 
Detection can also be achieved through lyrics, by exploiting statistical and semantic patterns of AI-generated text~\cite{frohmann2025double, frohmann2025ismir}, offering greater robustness to audio perturbations, but with limited performance. 
Beyond detection, complementary work addresses transparency and data replication, using notably audio similarity metrics to assess whether AI-generated music reproduces training content~\cite{batlleroca2023transparency,batlleroca2024replication}.

\textbf{Datasets:}
Several benchmarks have been proposed for AI music detection.
SONICS~\cite{rahman2024sonics} contains over 97k songs from Suno, Udio, and YouTube. While SONICS is a hybrid dataset at the prompt level with real or AI-generated lyrics/music styles, it does not mix human-recorded and AI-generated stems at the audio level. 
FakeMusicCaps~\cite{comanducci2025fakemusiccaps} targets both detection and attribution using five open-source text-to-music models.
ArtifactBench~\cite{oh2026artifactnet} spans 22 generators for codec-aware generalization evaluation, and M6~\cite{li2024m6} covers multiple generators, domains, languages, genres, and instruments.
AI-OpenBMAT~\cite{lopez2026ai} addresses broadcast scenarios with short, speech-masked excerpts, reporting F1-scores below 60\% when the SNR of the fake signal is low. This setting is related to ours, as AI-generated stems are mixed with human-recorded sources, but, as opposed to our study, the human-recorded part is considered as noise preventing detection. 
None of those datasets include \textit{hybrid} music that mixes AI-generated stems with real stems.

\textbf{Generated Singing Voice Detection in Music Mixtures:} 
The task of detecting whether vocals in a music track are real or AI-generated is the most closely related to ours. SingFake~\cite{zang2024singfake} addresses this by training several anti-spoofing countermeasure systems, either on source-separated vocals or full mixtures, finding that the benefit of source separation prior to detection is architecture-dependent, with no consistent conclusion across systems. Similarly, cloned singer identification shows that mixture-based models degrade substantially more than vocal-based ones when exposed to cloned voices~\cite{desblancs2024cloned}. A key distinction from both works is that our task is more general: rather than targeting just vocal authenticity, we perform \textit{stem-level} detection for potentially any component of the mix.

\section{Hybrid music dataset}\label{sec:dataset}

To our knowledge, no dataset currently exists for hybrid music containing musically coherent stems from real and AI-generated sources.
While building a realistic database in such settings from commercial AI generation services would require a dedicated work (left for future iterations), we decide here to synthesize hybrid mixture through a neural autoencoding of stems taken from a music source separation dataset, following the method in \cite{afchar2025fourier,afchar2025detecting}.

We use MUSDB18-HQ~\cite{musDB2019} source separation dataset providing multi-stems for 150 tracks, downmixed to isolated vocals $v$ and accompaniment $a$.
In order to synthesize the AI generation of a stem, we apply \textit{encodec} autoencoding with 24 kbps at 48 kHz configuration\cite{Defossez2022HighFNEncodec}.
This follows the framework introduced in~\cite{afchar2025fourier}, on which we base our work, and that reported robust detection performance for 3 reference neural codecs, as well as 3 versions of Suno and one of Udio.
We thus hypothesize that our experimental work and findings based on a single decoder should provide a robust base framework for extending the experiments to more realistic hybrid mixtures.

From both the original (with subscript $r$ for \emph{real}) and the autoencoded ($g$ for \emph{generated}) stems of a given track, we construct four types of coherent mixes:

\begin{itemize}
    \item \textbf{Hybrid mix with generated vocals}: encoded vocals with non-encoded accompaniment, noted $v_g + a_r$.
    \item \textbf{Hybrid mix with generated accompaniment}: non-encoded vocals with encoded accompaniment, $v_r + a_g$. 
    \item \textbf{Fully generated mix}: both stems encoded, $v_g + a_g$. 
    \item \textbf{Genuine mix}: both stems non-encoded, $v_r + a_r$.
\end{itemize}

The mixes are constructed linearly with different mixing gains.
In the remainder of this paper, when evaluating stem-specific detectors, we define as \textit{target gain} in $\{-12, -6, 0, +6, +12\}$\,dB, the gain applied to the target stem (\textit{e.g} the vocals for an AI vocals detection task) while the other stem's gain is set to 0 dB (\textit{e.g} the accompaniment).
Finally, a peak normalization to unit amplitude is applied to the obtained mixture. 
More complex mixing scenarios, including mastering, are left for future work. 
The full dataset comprises $150$ mixtures $\times 4$ mix types $\times 5$ target gains $= 3{,}000$ mixtures.

\begin{table}
\small
  \centering
  \begin{tabular}{|c|c|c|c|c|c|c|c|}
    \hline
    target gain (dB) & -12 & -6 & 0 & 6 & 12 \\
    \hline
    $v_g+a_r$ & 12.0 & 31.33 & 55.33 & 78.67 & 88.67 \\
    %\hline
    $v_r+a_g$ & 97.33 & 98.67 & 98.67 & 99.33 & 100.0 \\
    \hline
  \end{tabular}
  \caption{Proportion (in \%) of tracks detected as fully generated when the reference detector is directly applied to hybrid mixtures, for different mixing gains.}
  \label{tab:global_mix_hybrid_detection}
\end{table}

\section{Fully generated AI music detection}
\label{sec:fakemix_globaldetector}

This section details the training and evaluation of fully generated AI music detectors that will be used in Sections \ref{sec:naive_source_separation_detector} and \ref{sec:hybrid_detection_approach} for building stem-specific detectors.

\subsection{Reference detector}
\label{sec:ref_detector}

In this preliminary experiment, we assess the limits of an artifact-based AI detector trained from real and fully generated music, when applied to hybrid music. 
Thus, here the task is a global "generated \textit{vs.} real" binary classification of a track, with no detection at the stem level.

\textbf{Training on fully generated music:}
As our reference AI detector, we reproduce the system from~\cite{afchar2025fourier}.
We train this detector from scratch using the raw audio files from the Free Music Archive (FMA) medium dataset~\cite{fma_dataset} as the negative class (real), and the same files autoencoded with \textit{encodec}~\cite{Defossez2022HighFNEncodec} as the positive class (generated). 
We use a random train/test split of 70/30. 
Both real and generated versions of a track are distributed in the same set.
On the test set, we obtain a True Positive Rate (TPR) of $99.79\%$ and a False Positive Rate (FPR) of $0\%$, which is consistent with~\cite{afchar2025fourier}. Note that this task is performed on mixes only: the detector is a single binary classifier that discriminates fully real \textit{vs.} fully generated tracks, and does not provide any information about the generation at the stem level. 
Interestingly, when testing on the (unseen) dataset from Section~\ref{sec:dataset} on $v_r+a_r$ and $v_g+a_g$ scenarios, the detector achieves perfect accuracy ($100\%$ TPR and $0\%$ FPR).
%\newline
%- Train: TPR=0.999285, FPR=0  
%- Test:  TPR=0.997916, FPR=0

\textbf{Testing on hybrid mixtures:}
Here we apply the reference AI music detector to hybrid tracks, \textit{i.e.} to $v_r+a_g$ and $v_g+a_r$ subsets from the dataset of Section~\ref{sec:dataset}. 
The results, presented in \tabref{tab:global_mix_hybrid_detection}, show that the artifacts of the generated stem persist in the hybrid mixtures and that their detection is influenced by the target gain and the nature of the stem source. % (gain of the generated stem here). %, which in turn relates to the energy ratio and spectral content differences between both stems. 
We observe an asymmetry between vocals and accompaniment with more detection in the case $v_r+a_g$. This is probably because accompaniments that include percussion are usually dominant in high frequency bands where the detection happens (5-16kHz). 
Vocals also tend to be sparser over time and might be a minor contributor to the average profile computed by the detection system~\cite{afchar2025fourier}.
Therefore, while a plain binary detector is unable to distinguish between hybrid and fully generated mixes, this preliminary result suggests that useful detection artifacts are preserved during the mixing operation.

\subsection{From global to local AI detection}
\label{sec:local_detection}
We here adapt the fully generated AI detector, as presented above, from a global (full track prediction) to a local scale (short segment prediction).
This will allow for a finer temporal granularity essential to our proposed approach, as described in the following sections.
We thus re-train detectors on short audio chunks of $1$, $2$, $5$ and $10$ seconds from the same FMA-based dataset, keeping the same experimental conditions.
\tabref{tab:chunk_mix_detect} reports the AI chunk detection performance for all configurations.
%It corresponds to fakeprint averaged from respectively 4, 9, 25 and 160 STFT frames) instead of full tracks.
Performance remains high with only limited degradations for very short segments where only a fraction of audio is provided. 

\begin{table}
\small
  \centering
  \begin{tabular}{|l|c|c|c|c|c|c|c|}
    \hline
    Chunk dur. (s) & 10 & 5 & 2 & 1 \\
    \hline
    TPR (\%) & 99.733 & 99.693 & 99.139 & 96.860 \\
    %\hline
    FPR (\%) & 0.025 & 0.081 & 0.516 & 2.015 \\
    \hline
    %track TPR (\%) & 99.792 & 99.792 & 99.833 & 99.833 & 99.792 \\
    %track FPR (\%) & 0.000  & 0.000 & 0.000 & 0.000 & 0.000 \\
    %\hline
  \end{tabular}
  \caption{Fully AI-generated chunk prediction performance as a function of the audio chunk duration (in seconds).}
  \label{tab:chunk_mix_detect}
\end{table}

\section{Detection of AI-generated stems: A naive approach}
\label{sec:naive_source_separation_detector}

We now turn to the main task addressed in this work, namely AI-generated stem detection, and introduce a method that provides a prediction score for each stem (vocals or accompaniment) indicating whether or not it was generated.
This section explores a naive baseline solution to this task that consists of performing Music Source Separation (MSS) and then applying a plain AI music detector independently to each stem.

\subsection{Baseline: separate then detect}
Here, we use the dataset presented in Section \ref{sec:dataset}, considering the four mixing scenarios.
We first apply \textit{ht-demucs}~\cite{rouard2022hybrid} to perform 2-stems MSS on the mixtures, and then apply the global reference detector from Section \ref{sec:ref_detector} independently on each isolated stem.

%Demucs Note: clip mode=clamp to avoid messing with the estimated SNR in the following experiments, and other params as default)

The TPR and FPR for both generated vocals and accompaniment detection tasks are reported as blue bars in \figref{fig:hybrid_detection_performance} for the different target gains. 
Overall, the performance is quite poor, especially for the detection of generated vocals. 
Once again, the performance varies monotonically with respect to the mixing gain of the target stem, with very high FPRs for both vocals and accompaniment detection at lower target gains.

\begin{figure}
  \centering
  \includegraphics[width=\linewidth, trim={0 3mm 0 2mm}, clip]{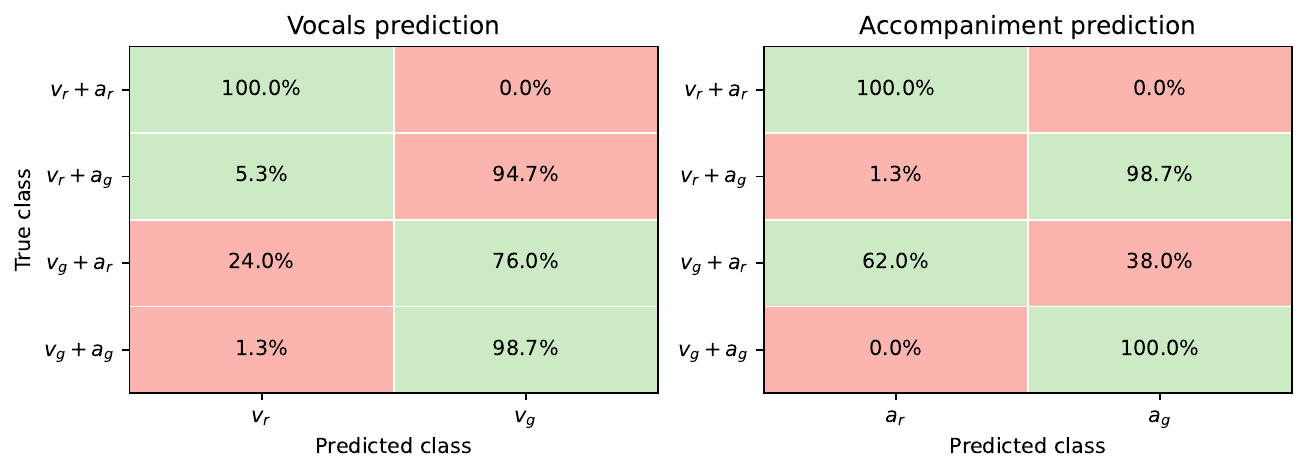}
  \caption{Confusion matrices for the naive stem detectors for all mixing cases with target gain $0$dB. Correct (resp. False) detection cases are colored in green (resp. red).}
  \label{fig:confusion_naive}
\end{figure}

Confusion matrices, represented in~\figref{fig:confusion_naive}, highlight that for the hybrid case $v_r+a_g$, the real vocals are almost always predicted as generated with a FPR of $94.7\%$ (left matrix). 
This suggests that artifacts from the generated accompaniment may have spread in the vocal stem through MSS. 
Thus, this naive system is unable to discriminate between $v_r+a_g$ and $v_g+a_g$. 
The discrimination between $v_g+a_r$ and $v_g+a_g$ is also weak, with a FPR of $38\%$ for the detection of accompaniment in the hybrid case $v_g+a_r$.  

\subsection{Qualitative analysis from the local AI detection}

We hypothesize that the naive approach performs poorly due to the MSS inability to accurately transfer the artifacts to the generated stem only, and propose a qualitative analysis based on the local AI-generated music detector from Section \ref{sec:local_detection}.
Figures \ref{fig:track_local_predictions} (a) and (b) illustrate the case of a hybrid track with generated vocals ($v_g+a_r$). 
Interestingly, local detection scores from the mix (blue), vocals (green) and accompaniment (orange) from subfigure (b) are highly correlated, while we would expect here that the accompaniment score remains close to 0 along the whole excerpt.
This tends to confirm our assumption that generation artifacts are spread over all separated sources.

We must note here that the MSS system can also introduce its own artifacts, potentially of a similar nature to those expected from neural codecs. 
Indeed, \textit{ht-demucs}~\cite{rouard2022hybrid} is based on an encoder/decoder-like architecture, involving upsampling from a latent space with transposed convolution layers.
While such artifacts are likely to appear, it seems from Figure~\ref{fig:track_local_predictions}(b) that this did not perturb our experiments: both local detection scores from extracted vocal and accompaniment stems return to 0 after $~177$ $s$, \textit{i.e.} when the generated stem is inactive.

While we could consider several, probably limited, improvements to this naive baseline (\textit{e.g.} testing different MSS architectures, retraining a MSS model from hybrid mixtures, or more simply retraining stem-specific detectors from separated sources), in the following section we propose an alternative approach that helps circumventing the exposed weaknesses of this sequential approach.

\section{Proposed approach}\label{sec:hybrid_detection_approach}

Building on preliminary experiments described in Section~\ref{sec:naive_source_separation_detector}, we propose here a joint/parallel approach that combines the prediction of the local mix detector with the information about the relative time-frequency energy of each stem.

\subsection{Formalizing the task}\label{sec:task_form}

Independently for each stem, we design a binary classifier that predicts from the audio mixture, at the chunk level (indexed by $t$), whether the considered stem was generated or not. 
The goal is to compute an estimate of $p(v_g | f_t, \{ \textrm{SNR}_t^b, b \in B\})$ (resp. $p(a_g | f_t, \{ \textrm{SNR}_t^b, b \in B\})$), the posterior probability that the vocal (resp. the accompaniment) stem is generated, given the prediction of the local AI mix detector $f_t$, and the ratio of energy between the target and complementary stems $\textrm{SNR}_t^b$ for each band $b$ in the set of all frequency bands considered $B$.
 
$f_t$ is obtained directly as the output of the local detector of Section \ref{sec:local_detection} applied to the input mix chunk $t$. 
The energy quantities used to compute $\textrm{SNR}_t^b$ are defined as the sum of the squared magnitudes of the short-time Fourier transform over the considered stems at chunk $t$ and frequency band $b$. 
We consider two scenarios for estimating $\textrm{SNR}_t^b$: 

\begin{itemize}

    \item \textbf{MSS SNR}, where we apply \textit{ht-demucs} on the mix and use the separated stems to compute the source energies.
    
    \item \textbf{Oracle SNR}, where the energy of the sources is directly computed from the available reference stems of the dataset, therefore bypassing the potential degradations and artifacts introduced by MSS.
    
\end{itemize}

In the following, we train a model at the chunk level $t$ (thus considering the independence of chunks) for $p(v_g | f_t, \{ \textrm{SNR}_t^b, b \in B\})$ and $p(a_g | f_t, \{ \textrm{SNR}_t^b, b \in B\})$.
Then, in order to get a prediction at the level of the full track, we compute the average over time of these probabilities and detect the stem as generated if it is above $0.5$, and as not generated otherwise.

\subsection{Experiments}\label{sec:hybrid_exp}

We use $2$-second-long chunks with $90\%$ overlap between chunks. 
We drop chunks with very low energy, considered as silence, as the corresponding $f_t$ and $\textrm{SNR}_t$ are unreliable in both train and test sets.  
This pre-processing removes on average $\sim 5\%$ of chunks per track, essentially trimming leading and trailing silences on the mix.
% quick and dirty implem: for each $track data_df['mix_energy'] / data_df['mix_max_energy']) <= 2/100$

We train a simple multilayer perceptron model (5 hidden layers, 32 linear units each, ReLu activation, sigmoid on output layer) for AI-generated vocals detection and a second independent one for AI-generated accompaniment detection. 
We use binary cross-entropy as loss function. %Other hyperparameters details are given in the code.
$f_t$ and $\textrm{SNR}_t^b$ input features are simply stacked together (total dimension of $B+1$) as input to the model.
%- simple NN (5 hidden layers, 32 linear units each, ReLu, output layer 1 linear unit sigmoid), binary cross entropy, no fine tuning tested on the model architecture
%- optim adam, LR=1e-3, batch 1024 shuffle, 50 epochs (all samples seen in 1 epoch) max, stopping after 5 epochs with no improvement on train loss
We run 10 cross-validation 70/30 splits, where the split is applied at the track level (\textit{i.e.} all mixing variants of a given track are distributed into the same set).
%- split 70/30 at the track level (for each run training on 105 tracks x 5 vocals mixing gains x 4 fakeness conditions, testing on the remaining 45 tracks x 5 x 4)

For the SNR bands, we use two configurations: 
\begin{itemize}
    \item 16 bands with 1 kHz width over the range 0-16 kHz, used for reporting performance in \ref{sec:hybrid_results}.
    \item A single band 5-16 kHz (which corresponds to the band used in~\cite{afchar2025fourier} for computing the detector's input features), only used for the qualitative analysis in Section \ref{sec:qualitative_analysis}. 
\end{itemize}
%2 chunk durations tested: 5 sec, and 2 sec (best)  => we stick to 2sec.

\begin{figure*}
    \centering
    \includegraphics[width=\linewidth, trim={0 2mm 0 2mm}, clip]{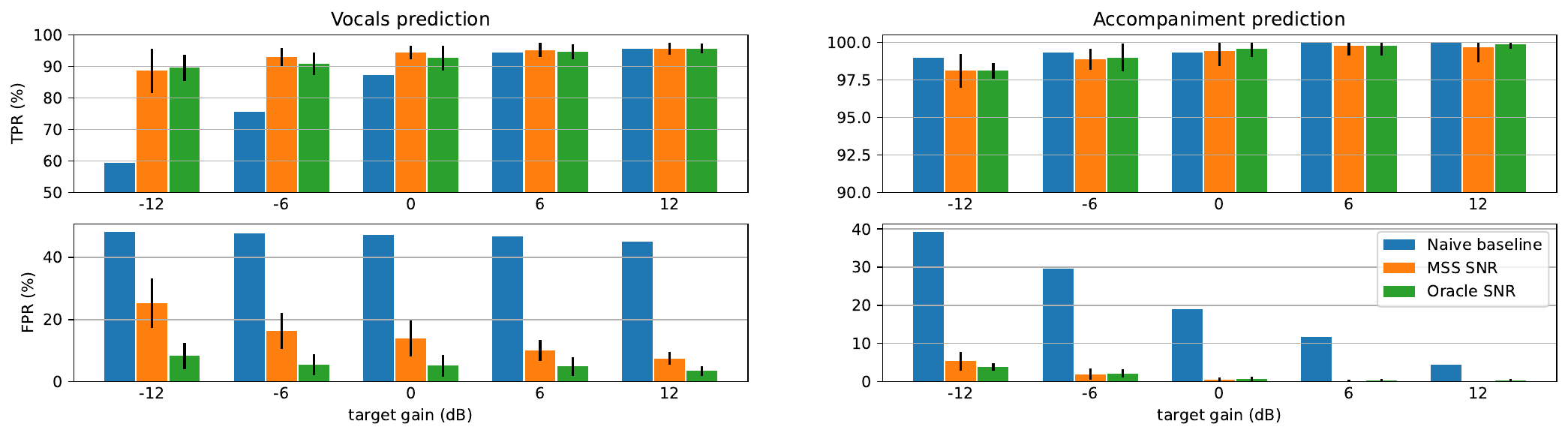}
    \caption{TPR (top) and FPR (bottom) metrics for vocals (left) and accompaniment (right) AI-generated stem detections grouped over all mixing configurations, and as a function of the target gain (x-axis).}
    \label{fig:hybrid_detection_performance}
\end{figure*}

\subsection{Evaluation}\label{sec:hybrid_results}

TPRs and FPRs are reported in~\figref{fig:hybrid_detection_performance}. 
We can see that the performance in the Oracle scenario (green bars) is quite high. 
As expected again, we observe that both metrics degrade as the target gain decreases: detection becomes more difficult when the generated stem has lower energy.
We also observe that performance is stem-dependent and that it remains much easier to detect generated accompaniment than generated vocals.

In the MSS scenario, general trends are similar (orange bars). 
TPRs are comparable to the Oracle configuration, but we notice a significant FPR degradation for the prediction of generated vocals, especially at low vocals gain.
In this realistic configuration, however, performance remains substantially better than the naive method (blue bars).

\begin{figure}
  \centering
  \includegraphics[width=\linewidth, trim={0 3mm 0 2mm}, clip]{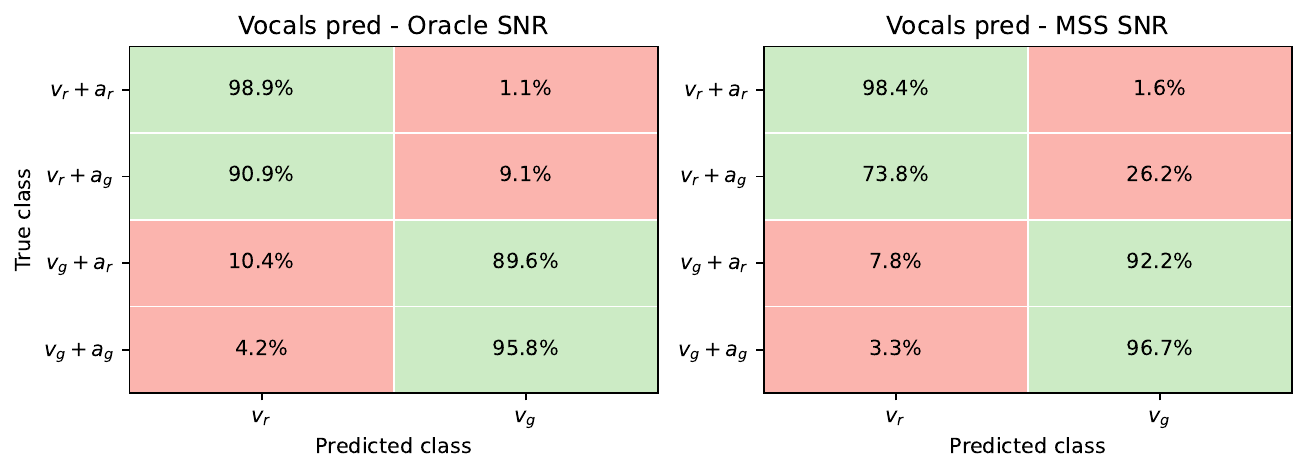}
  \caption{Confusion matrices for the proposed AI-generated vocals detector for a target gain of $0$dB. Influence of MSS (right) over Oracle (left) SNR estimator.}
  \label{fig:vocal_confusion_matrix}
\end{figure}

For a more detailed analysis of errors from the the vocals detector, we display in \figref{fig:vocal_confusion_matrix} the confusion matrices against the $4$ possible mixing classes at target gain 0dB, for both SNR scenarios.
We notice that, once again, the most difficult case is the discrimination between the hybrid mix $v_r+a_g$ and the fully generated mix $v_g+a_g$.
The realistic system using SNR estimated from MSS achieves the most degraded performance when compared to the Oracle SNR configuration. 
However, as opposed to the naive method (Fig. \ref{fig:confusion_naive}, left matrix), our system shows a stronger ability to discriminate between the hybrid case $v_r+a_g$ and the fully generated case $v_g+a_g$. 
Even in the realistic scenario using MSS-based SNR, for hybrid mixes $v_r+a_g$ the FPR drops from $94.7\%$ for the naive approach to $26.2\%$ for the proposed approach.

We must acknowledge here that Demucs was trained on MUSDB18-HQ. 
This could possibly result in better local SNR estimates than on data not seen during Demucs training, and in turn lead to overoptimistic performance. 
This would require retraining Demucs with other data, or collecting another dataset, which is left for future work.

\subsection{Qualitative analysis}
\label{sec:qualitative_analysis}

\begin{figure}
  \centering
  \includegraphics[width=\linewidth]{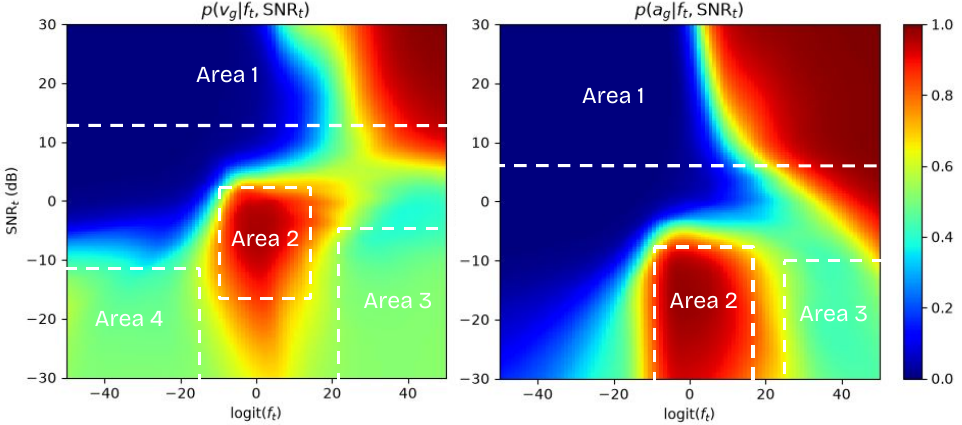}
  \caption{Trained posteriors for the generated vocals (left) and accompaniment (right) detection models.
  X-axis: logit-transformed fake score; how strongly the local detector predicts the mix as fake.
  Y-axis: SNR in dB; how loud the target stem is relative to the complementary stem.}
  \label{fig:posteriors}
\end{figure}

In \figref{fig:posteriors}, we display the trained model posteriors $p(v_g | f_t, \textrm{SNR}_t^b)$ and $p(a_g | f_t, \textrm{SNR}_t^b)$ for the simple case where there is a single SNR band from 5\,kHz to 16\,kHz. 
The x-axis shows the logit-transformed local mix detection score $\text{logit}(f_t) = \log(f_t / (1-f_t))$, mapping the output from $[0,1]$ to $]-\infty, +\infty[$, where negative (resp. positive) values indicate that the local mix detector predicts the segment as real (resp. generated). 
The y-axis represents the target stem SNR in dB. 
Although the decision frontiers are stem-dependent, a few consistent areas can be interpreted:
\vspace{-1.2em}
\paragraph*{Area 1:} for $\textrm{SNR} \gg 0$, the target stem (noted $s \in \{v, a\}$) is the dominant stem, then the prediction of generation for this stem tends towards the prediction of the mix: $p(s_g| f_t, \textrm{SNR}_t^b) \xrightarrow[\textrm{SNR}_t^b \rightarrow +\infty]{} f_t$.

For $\textrm{SNR} \ll 0$, \emph{i.e.} when the complementary stem is dominant, we observe three main areas: 
\vspace{-1.2em}
\paragraph*{Area 2:} for middle prediction scores on the mix ($0 \ll f_t \ll 1$), \emph{i.e.} when the mix detector is less confident, the target stem which has lower energy is likely generated.
Our interpretation is the following: for middle prediction scores in this specific 2-stem mixture scenario, at least one and only one stem has to be generated. 
Indeed, if both were real (resp. generated) the local mix detection score would tend towards 0 (resp. 1).
The complementary stem is here dominant and thus less likely to be generated; if it was, it would trigger a greater prediction score towards 1.
\vspace{-1.2em}
\paragraph*{Area 3:} $f_t \rightarrow 1$, the detector's score on the mix is very high. 
It is then likely that either the mix is fully generated, or the complementary dominant stem only is generated. 
The target stem then has equal chances to be generated or real, hence $p(s_g|f_t, \textrm{SNR}_t^b) \sim 0.5$.
Note that this value is due to the balance of the 4 classes of mix in our dataset, and can be interpreted as an uninformative prior.
\vspace{-1.2em}
\paragraph*{Area 4:}$f_t \rightarrow 0$, the detector on the mix has a very low prediction value. This means either that both stems are real, or that the energy of the target stem is so low that its generation artifacts disappear in the mix. Then, the target stem $s$ can either be generated or real, and $p(s_g|f_t, \textrm{SNR}_t^b) \sim 0.5$. 
This phenomenon appears mostly for vocals, though. For accompaniment, $p(a_g|f_t,\textrm{SNR}_t^b)$ drops to $0$, which is likely due to the fact that mixing with a vocals stem, even with high energy, is insufficient to hide generation artifacts (possibly from drums sitting in the hi-frequency range). 
Again, the value $0.5$ can still be interpreted as an uninformative prior.
    
This analysis gives an overview of the informative zones for differentiating hybrid mixes. Some of them are intuitive: in an area where a stem is strongly dominant, the prediction on the mix transfers to the stem. 
However, some are counterintuitive, as one may consider a low energy ratio for the target stem to result in an uninformative zone, when such a ratio can actually be very informative for middle-valued detection scores. 
Additional tests suggested that these properties remain valid in the case of $\textrm{SNR}_t^b$ estimated for multiple bands $b$.

These conclusions also describe hard samples: these are tracks for which the target stem SNR remains low while the prediction score $f_t$ remains extreme. 
On such samples, the posterior of the low-energy stem is uninformative and does not enable the discrimination between a fully generated mix and a hybrid mix where only one stem with high energy was generated.
The larger zone for hard samples in $p(v_g | f_t, \textrm{SNR}_t^b)$ may explain the loss of performance between vocals and accompaniment stem detection.

\begin{figure}
\centering
  \includegraphics[width=1\linewidth, trim={3mm 2mm 3mm 2mm}, clip]{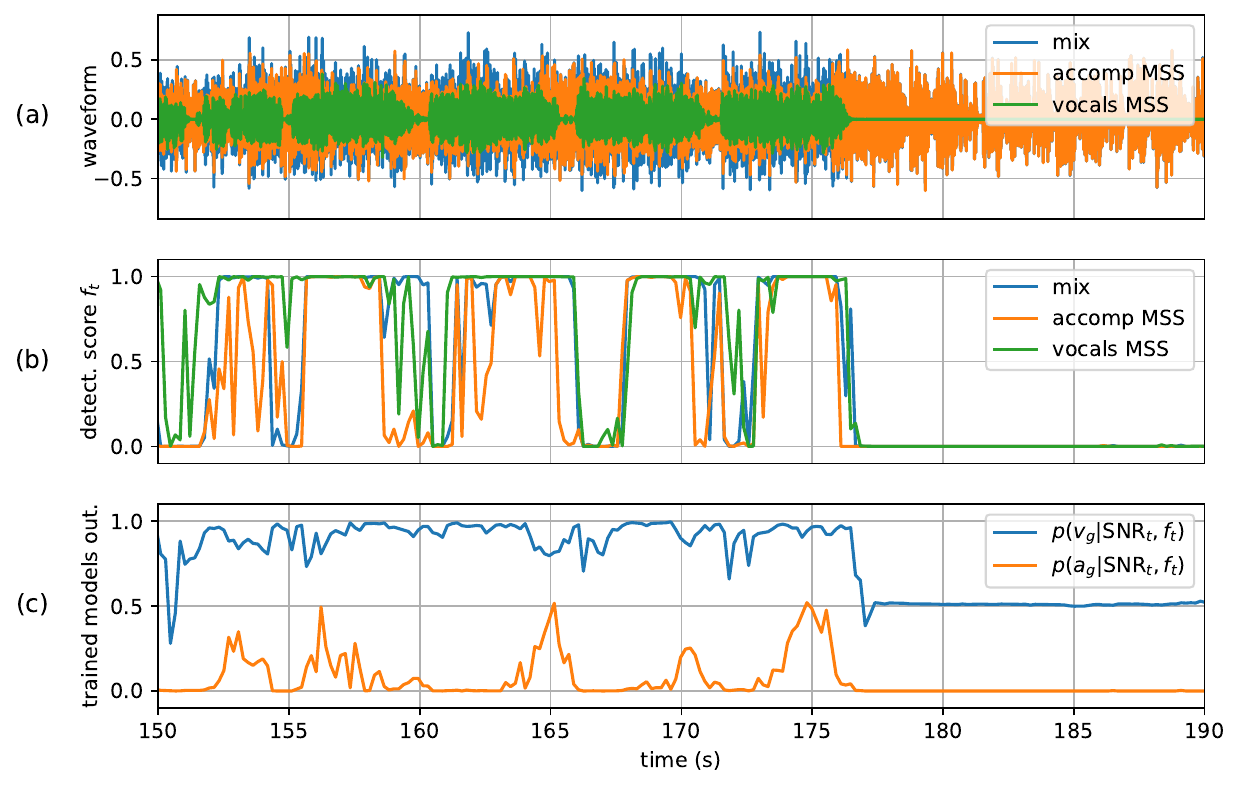}
  \caption{
  Illustration of the AI-generated stem detection for a hybrid track with generated vocals ($v_g + a_r$) at 0dB.
  (a) Waveforms of mix, and sources obtained through MSS.
  (b) Local detection on 2-sec chunks, run independently on the mix and on both MSS extracted sources, according to the naive approach from Section~\ref{sec:naive_source_separation_detector}, revealing that artifacts from the vocals have spread to the separated accompaniment.
  (c) Output of the local stem-specific detectors $p(v_g | f_t, \{\textrm{SNR}_t^b, b \in B\})$ and $p(a_g | f_t, \{\textrm{SNR}_t^b, b \in B\})$ from Section~\ref{sec:hybrid_detection_approach}.}
  \label{fig:track_local_predictions}
\end{figure}

\figref{fig:track_local_predictions}(c) depicts the local estimation of $p(v_g | f_t, \{\textrm{SNR}_t^b, b \in B\})$ and $p(a_g | f_t, \{\textrm{SNR}_t^b, b \in B\})$ on a hybrid track $v_g+a_r$ (generated vocals and real accompaniment). 
The estimation here is consistently correct with a high generation probability for vocals and a low generation probability for accompaniment. 
The system becomes less confident for the accompaniment when the vocal activity decreases. 
However, after $177 s$, when only the accompaniment is active, the probability for fake vocals approaches $0.5$ (uninformative chunks, in accordance with leftmost \figref{fig:posteriors}, Area 4), and the probability for fake accompaniment drops close to $0$ (low score on the mix prediction and high SNR for this stem, \textit{i.e.} leftmost part of Area 1 from \figref{fig:posteriors} right, so very confident prediction).
This suggests that local detections could be used as support statistics that can help in interpreting a detection decision taken at the global track-level.

\section{Conclusion}

In this paper, we presented a new method for independently detecting whether stems from a mix are AI-generated. 
We first highlighted the limitations of existing state-of-the-art detection in hybrid music settings. 
After showing that a naive approach that applies detection on the separated stems fails, we showed that a method that combines local prediction of generation on the mix and time-frequency activity estimates of the stems makes it possible to achieve promising performance for this new task. 
For a stem that dominates in the frequency band where the detection is performed (accompaniment in our experiments), we obtain very good detection performance and are able to discriminate the fully generated case from the generated-accompaniment-only case. 
For a stem that has less energy in the frequency band of the detection (vocals in our experiments), the performance is lower but still far above the naive baseline. 
Furthermore, the proposed local detection provides an interpretable way to understand global prediction.

Our approach has several limits, though.
First, in practical use cases, more than two stems might be mixed, and each may have been recorded or generated by a different model. 
In a generic multi-stem context, combinatorics will play against the method (SNR must be defined w.r.t. all other stems). 
Modeling the temporal relationship between the audio chunks might help in this context.
Moreover, MSS-based SNR estimation is computationally costly and could likely be replaced by simpler models that estimate the local energy of stems with a lower resolution.
Finally, the method could benefit from various improvements, such as learning a detection pattern for each stem, or learning local time-frequency mix detectors separately on the frequency bands used to compute SNRs. 
This would potentially favor mix detection scores with extreme values for which stem prediction appears to be easier.

\section{Ethics statement}

While AI-generation detectors can be useful tools for improving transparency on music streaming platforms and catalog monitoring in the whole music recording industry, we believe that they are not the unique answer to the spread of AI-generated music. 
Detectors are not universal, are prone to errors, are not robust to audio manipulation, and are designed with specific simplified scenarios in mind (until now, most papers only modeled the task as a simple binary classification). 
Consequently, we believe that these detectors must be used with care, always reminding these limitations.

We must acknowledge that 1) we cover the task of stem-level generation detection in a restricted scenario of vocals and accompaniment, which, while arguably common, does not cover all the possibilities of hybridation between AI-generated content and human-produced content. 2) The performance of our system is still modest, with a significant amount of false positives.
Consequently, we believe that the usage of such a system should consider the restriction of this scenario and would need careful tuning to mitigate false positives. 
Also, it should allow for recourse, especially in situations where it can have a negative financial impact on actual artists.

% For BibTeX users:
%\clearpage 
\bibliography{ISMIRtemplate}

\end{document}